\documentstyle[12pt]{article}
\newcommand{\be}{\begin{equation}}
\newcommand{\ee}{\end{equation}}
\newcommand{\ba}{\begin{eqnarray}}
\newcommand{\ea}{\end{eqnarray}}
\newcommand{\bastar}{\begin{eqnarray*}}
\newcommand{\eastar}{\end{eqnarray*}}
\newcommand{\half}{\frac{1}{2}}
\newcommand{\g}{{\bf \hat g}}

\newcommand{\X}{{\cal X}}
\newcommand{\om}{\omega_{ij}}
\newcommand{\Z}{{\cal Z}}

\newcommand{\cL}{{\cal L}}

\newcommand{\mlabel}[1]{\label{#1}}

\begin{document}
\begin{titlepage}
\begin{flushright}
UU-ITP 02/94 \\
HU-TFT-94-08 \\
hep-th/9402041
\end{flushright}

\vskip 0.7truecm

\begin{center}
{ \bf ON THE DUISTERMAAT-HECKMAN FORMULA \\ \vskip 0.2cm
AND INTEGRABLE MODELS$^{\dag}$  \\  }
\end{center}

\vskip 1.5cm

\begin{center}
{\bf T. K\"arki  ~~and~~ Antti J. Niemi$^{\star}$  } \\
\vskip 0.4cm
{\it Department of Theoretical Physics, Uppsala University
\\ P.O. Box 803, S-75108, Uppsala, Sweden$^{\star}$    \vskip 0.2cm
\vskip 1.0cm
and
\vskip 1.0cm
Research Institute for Theoretical Physics, Helsinki University \\
Siltavuorenpenger 20 C, FIN-00170 Helsinki, Finland}

\end{center}

\vskip 0.5cm

\

\vskip 2.7cm

\rm
\noindent
In this article we review the Duistermaat-Heckman integration
formula and the ensuing equivariant cohomology structure,
in the finite dimensional case. In particular, we discuss the
connection between equivariant cohomology and classical
integrability. We also explain how the integration formula is
derived, and explore some possible new directions that could
eventually yield novel integration formulas for nontrivial
integrable models.

\vfill

\begin{flushleft}
\rule{5.1 in}{.007 in}\\
$^{\dag}$ {\small based on presentation by A.N. at XXVIIth
International
Symposium on the Theory of Elementary Particles Wendisch-Rietz
(Germany)
September 7-11, 1993 \\}
$^{\star}$ {\small permanent address \\}
{\small E-mail addresses: \bf tkarki@phcu.helsinki.fi $~~~$
niemi@rhea.teorfys.uu.se \\}
\end{flushleft}

\end{titlepage}

\vfill\eject

\baselineskip 0.65cm

{\bf 1. Introduction}
\vskip 1.0cm

The Duistermaat-Heckman integration formula \cite{seme},\cite{duis}
and its
generalizations to path integrals have been discussed
extensively both in the Mathematics \cite{berl}-\cite{witt} and in
the Physics
\cite{omat}-\cite{gord} literature.
In this article, we shall review the original integration formula and
show how it can be derived. In particular we shall discuss the
equivariant cohomology structure which underlies the
integration formula, explore connections between equivariant
cohomology and classical integrability, and investigate how the known
integration formulas could be generalized.

We shall limit ourselves to
finite dimensional integrals, and we refer to the original articles
[5,6] where
 path integral generalizations have been discussed: The integration
formulas
for path integrals
can be obtained by generalizing the present
construction to the loop space. Even though there are some genuine
loop space intricacies that are absent in finite
dimensions, the insight gained in finite dimensions is in any case
quite
applicable also in the loop space.

\vskip 1.5cm

{\bf 2. Equivariance, Integrable Models And Localization}

\vskip 1.0cm

We shall consider a $2n$ dimensional compact phase space $M$, with
local coordinates $z^{i}$ and Poisson bracket
\be
\{z^{i} , z^{j} \} ~=~ \omega^{ij}(z)
\ee
Here $\omega^{ij}$ is the inverse matrix to the symplectic
two-form on $M$,
\be
\omega ~=~ \half \om dz^{i} dz^j
\ee
This is closed,
\be
d\omega ~=~ 0
\ee
so that locally we can introduce the one-form $\vartheta$  called
the symplectic potential such that
\be
\omega ~=~ d\vartheta
\ee

We are interested in the exact evaluation of the "classical"
partition
function
\be
\Z ~=~ \int \omega^n e^{-\beta H}
\label{fpf1}
\ee
where $H$ is some hamiltonian on $M$, and $\beta$ is a real parameter
(inverse temperature in classical statistical physics). The
integration
formula by Duistermaat and Heckman states, that if $H$ determines the
symplectic action of $U(1)$ on the phase space the
integral (\ref{fpf1}) localizes to the critical points of $H$,
\be
\Z~=~ \frac{1}{\beta^n}\sum\limits_{dH=0} { \sqrt{det||\om ||}\over
\sqrt{ det|| \partial_{ij}  H || } } ~\exp\{ -\beta H\}
\label{fdh}
\ee
In this article we shall explain how (\ref{fdh}) is derived. We shall
also
discuss some possible approaches to generalize this integration
formula to a
wider class of Hamiltonians.

\vskip 0.5cm

The integration formula (\ref{fdh}) is derived using
equivariant cohomology on the phase space $M$.  For this we consider
the  exterior algebra $\Omega(M)$ of $M$ and
introduce the contraction operator $i_\X$ with respect to a
general vector field $\X$. It is a nilpotent operator on the
exterior algebra $\Omega(M)$. We also introduce the equivariant
exterior derivative
\be
d_\X ~=~ d + \phi i_\X
\label{fed}
\ee
where $\phi$ is a real parameter, and the Lie derivative along $\X$
\be
\cL_\X ~=~ d_\X^2 ~=~ \phi (d i_\X + i_\X d)
\ee
On the subcomplex $\Omega_\X$ of $\X$-invariant
exterior forms
\be
\cL_\X \Omega_\X ~=~ 0
\ee
the exterior derivative (\ref{fed}) is then nilpotent and defines the
$\X$-equivariant subcomplex $\Omega_\X$ of $\Omega(M)$. The
corresponding
cohomology determines the  equivariant cohomology on $M$.

We shall assume that the action of $\X$ on $M$ is
symplectic,
\be
\cL_\X \omega ~=~ d i_\X \omega ~=~ 0
\label{fsa}
\ee
Provided the one-form $i_\X\omega$ is exact (for this
the triviality of $H^1(M,R)$ is sufficient), we can
introduce the corresponding Hamiltonian $H (z)$
\be
i_\X \omega ~=~ - d H
\label{fhe}
\ee
In local coordinates $z^i$ on $M$ this becomes
\be
\X ~=~ \omega^{ab}\partial_a H  \partial_b
\ee

\vskip 0.4cm
In order to use this formalism to derive the integration
formula (\ref{fdh}) and its generalizations, we realize the
various operations on the exterior algebra $\Omega(M)$
canonically. For this we introduce a canonical conjugate variable
$p_i$ for $z^i$, identify $dz^i$ with anticommuting $c^i$ and the
contraction operator on  $c^i$ with ${\bar c}_i$, with Poisson
brackets
\ba \{ p_i , z^j \} ~=~ \delta_i^j
\\
\{ \bar c_i , c^j \} ~=~ \delta_i^j
\ea
In terms of these variables  the exterior derivative, contraction and
Lie derivative can be realized by the Poisson bracket actions of
\be
d ~=~ p_i c^i
\label{fcr1}
\ee
\be
i_H ~=~ \X^i \bar c_i
\label{fcr2}
\ee
\be
\cL_H ~=~ \X^i p_i + c^i \partial_i \X^j \bar c_j
\label{fcr3}
\ee

\vskip 0.4cm
\noindent
Since
\be
d_H (\phi H + \omega ) ~=~ \phi ( dH + i_H \omega)
= 0
\ee
by (\ref{fhe}),  we conclude that $\phi H + \omega$ is an element of
$H^*(M)$
and determines an equivariant cohomology class.
This is an equivalence class consisting of elements in $\Omega(M)$
which are
linear combinations of zero- and two-forms that can be represented as
\be
\phi H + \omega + d_H \psi
\ee
where $\psi \in \Omega(M)$ satisfies
\be
\cL_H \psi ~=~ 0
\label{fof}
\ee
and is a linear combination of the form
\be
\psi ~=~ \psi_0 + \psi_1 +\psi_2 + ~ ... ~ + \psi_{2n}
\label{cmp}
\ee
where $\psi_k$ is a $k$-form on $\Omega(M)$. In particular, due to
linearity of
$\cL_H$ these $k$-forms also satisfy
\be
\cL_H \psi_k ~=~0
\label{liecmp}
\ee

\vskip 0.2cm
Suppose now that there exists a $\psi$ which
satisfies (\ref{fof}) and in addition $d_H \psi$ is a linear
combination of a
zero- and two-form. Denoting the zero-form by $K$ and the two-form by
$\Omega$,
we then have the relations ($\phi = 1$ for simpler notation)
\ba
i_H \psi_1 ~=~ K \\
d\psi_0 + i_H \psi_2 ~=~ 0 \\
d\psi_1 + i_H \psi_3 ~=~ \Omega \\
d\psi_2 + i_H \psi_4 ~=~ 0
\label{fpo}
\ea
$$
etc ...
$$
In particular, since
$$
d i_H \psi_3 ~=~ - i_H d \psi_3 ~=~ i_H i_H \psi_5 ~=~ 0
$$
by (\ref{liecmp}), we conclude that $\Omega$ is a {\it closed}
two-form (but
not necessarily nondegenerate). Furthermore, since
\be
dK ~=~ di_H \psi_1 ~=~ - i_H d\psi_1 ~=~ - i_H \Omega
\ee
we get
\be
\Omega^{ab} \partial_b K ~=~ \omega^{ab}\partial_b H
\label{fbh}
\ee
so that the classical equations of motion for the two Hamiltonian
systems $(H,\omega)$ and $(K,\Omega)$ coincide. As a consequence
these two systems determine a {\it bi-Hamiltonian structure}. (Here
we assume
that $\Omega_{ab}$ is nondegenerate on $M$ except possibly on
submanifolds of
$M$ which have a co-dimension larger than or equal to two. On these
submanifolds, the Hamiltonian $K$ must then vanish
to keep the equations of motion non-singular.)

On the other hand, if we have a bi-Hamiltonian structure
so that (\ref{fbh}) holds we get
\be
d_H (\phi K + \Omega) ~=~ 0
\label{fko1}
\ee
that is, $\phi K+\Omega$ is equivariantly closed with respect to
$d_H$. We can
then apply an equivariant form of Poincare's lemma to conclude that
there exist
a (locally defined) form $\psi$ on $\Omega(M)$ such that
\be
\phi K + \Omega ~=~ d_H \psi
\label{fko2}
\ee
Indeed, for an {\it integrable} model this can be easily solved at
least
locally: If $(H,\omega)$ and $(K,\Omega)$ is an integrable
bi-Hamiltonian pair,
we can introduce action-angle variables $(I_i , \theta_i)$ (almost)
everywhere
on $M$ such that both $H$ and $K$ depend only on the action
variables, $H =
H(I)$ and $K = K(I)$. We assume that these variables are selected so
that
$\Omega$ admits the Darboux form
\be
\Omega ~=~ \sum\limits_i dI_i \wedge d\theta_i
\ee
while $\omega_{ab}$ remains a nontrivial function of
$(I_i,\theta_i)$.
The corresponding symplectic potential for $\Omega$ is
\be
\vartheta ~=~ \sum\limits_i I_i d\theta_i + dF
\ee
where $F$ is some function.

In the complement of the critical point set of $K$ we have an action
variable
$I_k$ such that
$$
{ \partial K(I) \over \partial I_k } ~\not=~ 0
$$
With $\theta_k$ the corresponding angle variable, we introduce
the following function on $M$
\be
W(I,\theta) ~=~ \theta_k \cdot \left( {\partial K \over
\partial I_k }\right)^{-1}
\ee
We then consider (\ref{fko2}) ($\phi = 1$ here for simplicity):
\be
K + \Omega ~=~ (d + i_H) (\vartheta + dF) ~=~
\omega^{ab}\partial_b H \vartheta_a + \omega^{ab} \partial_b H
\partial_a F +
\Omega
\ee
Using ( ), we get further
\be
=~ \sum\limits_i { \partial
K \over \partial I_i } I_i + \{ K , F \}  + \sum\limits_i dI_i
\wedge d\theta_i
\ee
Hence, if we select
\be
F(I,\theta) ~=~ W \cdot (K - \sum\limits_i { \partial K \over
\partial I_i }
I_i)
+ G(I)
\ee
with $G(I)$ an arbitrary function of the action variables, we
conclude that the
one-form $\psi = \vartheta + dF$ satisfies (\ref{fko2}),
\be
K + \Omega ~=~ d_H (\vartheta + dF)
\ee

The global existence of such a form $\psi$ is connected to the
nontriviality of
the equivariant cohomology associated with $d_H$.

\vskip 0.5cm

The preceding discussion suggests an intimate relationship between
equivariant
cohomology and the existence of bi-Hamiltonian structures, which
deserves
further investigation.

\vskip 0.4cm

We remind, that if the symplectic two-forms $\omega$ and $\Omega$ are
such that
the following rank-(1,1) tensor is {\it nontrivial},
\be
L ~=~ \Omega \omega^{-1} ~=~ \Omega_{ac}\omega^{cb} c^{a} p_b
\ee
we can establish the integrability of $(H,\omega)$ under a certain
condition: From (\ref{fpo}),
\be
\cL_H \Omega ~=~   (d  i_H + i_H d) d\psi ~=~   d i_H d \psi
\ee
and since $\psi$ satisfies (\ref{fof}),
we conclude that
\be
\cL_H \Omega ~=~ 0
\ee
Using (\ref{fsa}), we then get
\be
\cL_H L = \cL_H (\Omega \omega^{-1}) = (\cL_H \Omega) \omega^{-1}
+ \Omega (\cL_H \omega^{-1} ) = 0
\ee
or in components
\be
\X_H^c \partial_c L_a^b + L_c^a\partial_a \X_H^c
- L_a^c \partial_c \X_H^b = 0
\ee
This we can write as
\be
{dL_a^b \over dt } ~=~ L_a^c (\partial_c \X_H^b) ~-~ (\partial_a
\X_H^c) L_c^b
\ee
or
\be
{ dL \over dt } ~=~ [L , U ]
\ee
with $U_a^b = \partial_a\X_H^b$. This is the Lax equation and
$L, ~U$ is the Lax pair.

In order to prove integrability, we introduce
\be
Q_k ~=~ \frac{1}{k} Tr L^k
\label{fcc1}
\ee
and define the Nijenhuis tensor $N$, with components
\be
N_{ab}^c ~=~ L_a^d \partial_d L_b^c - L_b^d \partial_d L_a^c - L_d^c
(\partial_a L_b^d - \partial_b L_a^d)
\ee
Then,
$$
N_{ab}^c( L^{n-1})_c^b ~=~ (L_a^d \partial_d L_b^c - L_b^d \partial_d
L_a^c - L_d^c [ \partial_a L_b^d - \partial_b L_a^d]) (L^{n-1})_c^b
$$
$$
= ~ L_a^d (\partial_d L_b^c) (L^{n-1})_c^b - (L^n)_c^d (\partial_d
L_a^c) - (L^n)_d^b ( \partial_a L_b^d - \partial_b L_a^d)
$$
\be
=~ L_a^d(\partial_d L_b^c ) (L^{n-1})_c^b - (L^n)_d^b \partial_a
L_b^d
\ee
On the other hand,
$$
\partial_d Tr L^n ~=~ \sum_{m} Tr (L ... {\partial_d}
\!\!\stackrel{(m)}{L} ... L) ~=~ nTr(\partial_d L)L^{n-1}
{}~ = ~ n(\partial_d L_b^c) (L^{n-1})_c^b
$$
so that using
$$
\partial_d Q_n ~=~ (\partial_d L_b^c) (L^{n-1})_c^b
$$
we  conclude that
\be
N_{ab}^c( L^{n-1})_c^b ~=~ L_a^d \partial_dQ_n - \partial_a Q_{n+1}
\ee
As a consequence, {\it if the Nijenhuis tensor vanishes} we obtain
the
recursion relation
\be
L_a^d \partial_d Q_n ~=~ \partial_a Q_{n+1}
\ee
which we can also write as
\be
\omega^{ab}\partial_b Q_n ~=~ \Omega^{ab} Q_{n+1}
\ee
Furthermore, since
$$
\{ Q_i , Q_j \}_{\omega} = \omega^{ab}\partial_a Q_i \partial_b Q_j
= \Omega^{ab} \partial_a Q_i \partial_b Q_{j+1} = \omega^{ab}
\partial_a Q_{i-1} \partial_b Q_{j+1} = \{ Q_{i-1} , Q_{j+1}
\}_{\omega}
$$
and assuming $i > j$ and iterating $i-j$ times, we get
$$
\{ Q_{i} , Q_{j} \}_\omega ~=~ \{ Q_j , Q_i \}_\omega
$$
Hence
\be
\{ Q_i , Q_j \}_\omega ~=~ 0 ~~~~~~~~{\rm for ~ all ~~} i,j
\ee
and we have constructed $n$  quantities in
involution.  Furthermore, since
\be
Tr \{ L^k \frac{dL}{dt } \} ~=~ Tr \{ L^k [L,U] \} ~=~ 0
\ee
we conclude that these  quantities are conserved {\it i.e.}
commute with the Hamiltonian $H$. This establishes the integrability
of the
Hamiltonian system $(H,\omega)$, provided
(\ref{fcc1}) are complete {\it i.e.} the number of functionally
independent
(\ref{fcc1}) coincides with half the phase space dimension. However,
in general
there is no guarantee that these integrals are complete, and this
completeness
must be established independently.

\vskip 0.4cm
Consider now the integral
\be
\Z ~=~ \phi^n \int \omega^{n} e^{-\phi H}
\label{fpf2}
\ee
for some  Hamiltonian $H$ that admits a bi-Hamiltonian structure.  If
we introduce anticommuting variables $c^a$ we can write $\Z$ as
\be
\Z ~=~  (-)^n n ! \phi^{n} \int dz^i \sqrt{ det ||
\om || } e^{-\phi H}
{}~ =~ (-)^n n !  \phi^{n}   \int dz^i dc^i e^{-\phi H - \omega  }
\label{fef}
\ee
We assume that $\psi$ is a one-form such that
\be
\cL_H \psi ~=~ 0
\label{flp}
\ee
With $\lambda$ a real parameter we then argue that the
following one-parameter family of integrals
\be
\Z_\lambda ~=~ \int dz^i dc^i \exp\{ - \phi H - \omega - \lambda d_H
\psi \}
\label{fpf3}
\ee
does not depend on $\lambda$. This implies in particular,  that the
integral
(\ref{fpf2}) {\it only} depends on the equivalence class determined
by $(H,\omega)$ in the $d_H$ equivariant cohomology.

In order to prove this $\lambda$-independence, we consider an
infinitesimal variation $\lambda \to \lambda + \delta
\lambda$, and show that
\be
\Z_\lambda ~=~ \Z_{\lambda + \delta \lambda}
\ee
For this, we introduce the following infinitesimal change of
variables in (\ref{fpf3}):
\ba
z^i ~\to~ \tilde z^i ~=~ z^i ~+~ \delta z^i ~=~
z^i ~+~\delta \psi \cdot d_H z^i ~=~ z^i ~+~ \delta \psi c^i
\\
c^i ~\to~ \tilde c^i ~=~ c^i ~+~ \delta c^i ~=~ c^i~+~ \delta \psi
\cdot d_H c^i ~=~ c^i ~-~ \delta \psi \X^i
\label{fcuu}
\ea
with
\be
\delta \psi ~=~ \delta \lambda \cdot \psi
\ee
Since $\psi$ satisfies (\ref{flp}), the exponential in (\ref{fpf3})
is
invariant under
the change of variables (\ref{fcuu}). However, the Jacobian is
nontrivial:
$$
d\tilde z^i d \tilde c^i=Sdet\left(
\begin{array}{cc} {\partial \tilde z^i \over \partial z^i} & \tilde
z^i  \frac{\stackrel{\leftarrow}{\partial}}{\partial c^i}\\
 &  \\
{\partial \tilde c^i \over \partial z^i} &  {\partial
\tilde c^i \over \partial c^i}
\end{array} \right)dz^idc^i
$$
\vskip 0.4cm
$$
= ~ Sdet \left(
\begin{array}{cc} 1+\frac{\partial \delta \psi}{\partial z^i}c^i &
\star
\\
& \\
\star & 1 -\frac{\partial \delta \psi }{\partial c^i}\X^i
\end{array}
\right) dz^i dc^i
$$
\vskip 0.4cm
$$
=\left( 1~+~ Str \left( \begin{array}{cc}  \frac{\partial \delta
\psi}{\partial
z^i}c^i & \star \\
& \\
\star & -\frac{\partial \delta \psi}{\partial c^i}\X^i
\end{array}
\right)\right)dz^idc^i
$$
\vskip 0.4cm
$$
= ~ (1+ \frac{\partial \delta \psi}{\partial z^i}c^i+\frac{\partial
\delta
\psi}{\partial c^i}\X^i )dz ^i dc^i
$$
\vskip 0.2cm
$$
=(1-(c^i\frac{\partial}{\partial z^i}-\X^i
\frac{\partial}{\partial
c^i})\delta \psi)dz^i dc^i
$$
\vskip 0.2cm
$$
= ~(1-d_{H}\delta \psi )dz^i dc^i ~\sim~ \exp\{ -d_{H}(\delta
\psi)\}dz^idc^i ~=~ \exp\{ - \delta \lambda d_H \psi \}
$$
Hence
$$
\Z ~=~ \int dz^i dc^i \exp\{ - \phi H + \omega - \lambda d_H \psi -
d_H (\delta\psi) \}
$$
\be
=~ \int dz^i dc^i \exp\{ - \phi H+\omega - (\lambda + \delta
\lambda) d_H \psi
\} ~=~ Z_{\lambda + \delta \lambda}
\ee
and we have established that if the Hamiltonian system
$(H,\omega)$ admits a bi-Hamiltonian struture the classical partition
function (\ref{fpf2}) depends {\it only} on the equivalence class
that $(H,\omega)$ determines in the $d_H$ equivariant cohomology.

\vskip 1.5cm
{\bf 3. Duistermaat-Heckman Integration Formula}
\vskip 1.0cm

In order to construct examples  we consider a compact Lie group $G$
that acts
on $M$ by local diffeomorphisms which are generated by vector fields
$\X_a$,
$~a=1,...,m$. Their commutation relations defines a representation of
the Lie
algebra $\g$
of $G$,
\be [ \X_a , \X_b ] ~=~ f_{abc} \X_{c}
\ee
where $f_{abc}$ are the structure constants of $\g$.

We denote contraction {\it w.r.t.} the Lie algebra basis $\{ \X_{a}
\}$  by
$i_a$.  The corresponding Lie derivatives
\be
\cL_a ~=~ d i_a + i_a d
\ee
then generate the action of $G$ on the exterior algebra $\Omega(M)$
of
$M$,
\be
[\cL_a , \cL_b ] ~=~ f_{abc} \cL_c
\ee

We assume that the action of $G$ on $M$ is
symplectic,
\be
\cL_a \omega ~=~ d i_a \omega ~=~ 0 ~~~~~~~~{\rm for ~ all~} a
\ee
so that we can define the momentum map
\be
H_G: ~~~ M \mapsto \g^*
\ee
which gives a one-to-one correspondence between the vector
fields $\X_a$ and the components $T_a$ of the momentum map,
\be
H_G ~=~ \phi^a T_a
\label{fmm}
\ee
where $\{ \phi^a \}$ is a (symmetric) basis of $\g$.
{}From the Jacobi identity for $\g$ we then get the homomorphism
\be
[ \X_a , \X_b ] ~=~ f_{abc} \X_c  ~=~ \X_{ \{ T_a , T_b \} }
\ee
However, in general the Hamiltonian corresponding to the commutator
of
two generators may differ from the Poisson bracket of the
corresponding
Hamiltonians by a two-cocycle,
\be
\{ T_a , T_b \} ~=~  f_{abc} T_c + \kappa_{ab}
\ee
but here we shall assume that $\kappa_{ab}=0$.
\vskip 0.3cm
\noindent
We again introduce the canonical realization
(\ref{fcr1})-(\ref{fcr3}) of the
various operations.

\vskip 0.5cm

The simplest example of the present construction is the canonical
action of the circle $ G=  U(1) \sim S^1 $. This action is generated
by a vector field $\X$, the generator of the Lie-algebra $u(1)$ of
$U(1)$.  The corresponding momentum map (\ref{fmm})   is
\be
H_{U(1)} ~=~ \phi H
\label{fhu}
\ee
and
\be
\X^a ~=~ \omega^{ab}\partial_b H
\ee
and $\phi$ is the generator of the dual basis
of $u(1)$, a real parameter.

We wish to derive the integration formula (\ref{fdh}) for
the integral
\be
\Z ~=~ \phi^n \int \omega^n e^{-\phi H}
\ee
with $H$ the hamiltonian in (\ref{fhu}). For this we introduce the
corresponding equivariant exterior
derivative
\be
d_H ~=~ d + \phi i_H
\mlabel{sabelcartan}
\ee
and the Lie-derivative with respect to $\X$ is
$$
d_H^2 =  \phi (d i_H +  i_H d) =  \phi \cL_H \ .
$$
so that on the subcomplex $\Omega_{U(1)}$ of $U(1)$-invariant
exterior forms, $d_H$ is nilpotent and defines an exterior
differential operator.  The pertinent cohomology
coincides with the equivariant cohomology $H^*_{U(1)}(M)$ of
the  manifold $M$.

In order to derive (\ref{fdh}), we first need a one-form $\psi$ that
we can use in (\ref{fpf3}), to localize it onto (\ref{fdh}). For this
we first
observe that since the group $G \sim U(1)$ is compact, we may
construct a
metric tensor $g_{ab}$ on $M$ for which the canonical flow of $H$ is
an
isometry,
\be
\cL_H g ~=~ 0
\label{frm}
\ee
Such a metric is obtained by first selecting an arbitrary Riemannian
metric  $\tilde g$ on $M$, and averaging it over the group $G$ using
its
Haar measure. A converse  is also true:  Since $M$ is
compact the isometry group of $g$ must also be compact.

We select
\be
\psi ~=~ i_H g ~=~ g_{ij} \X^i c^j
\label{flf}
\ee
where $g$ is the Riemannian metric (\ref{frm}). As a consequence,
\be
\cL_H \psi ~=~ 0
\ee
and we obtain the bi-Hamiltonian structure with
\be
K ~=~ g_{ij} \X^i \X^j
\ee
\be
\Omega_{ij} ~=~ \partial_i ( g_{jk}\X^k) - \partial_j (g_{ik}\X^k)
\ee
and the integral
\be
\Z ~=~ \phi^n \int dz^i dc^i \exp\{ - \phi H - \omega +
\lambda(K+\Omega) \}
\ee
is independent of $\lambda$.

Explicitly,
\be
\Z ~=~ \phi^n \int dz^i dc^i \exp\{ - \phi H - \omega - \lambda
g_{ij}\X^i\X^j
- \half \lambda \cdot \Omega_{ij} c^i c^j \}
\label{fpl}
\ee
In the $\lambda \to \infty$  we can then use
\ba
\delta (z^i)=&\lim\limits_{\lambda \to \infty} \left( {\lambda
\over 2\pi } \right)^{\frac{n}{2}} \sqrt{\det || S_{ij} || } \cdot
e^{-\frac{\lambda}{2} z^i S_{ij} z^j}
\\
\delta (c^i) =&\lim\limits_{\lambda \to \infty} \lambda
^{-\frac{n}{2}} {1 \over \sqrt{\det || A_{ij} ||}} \cdot
e^{\frac{\lambda}{2}
c^i   A_{ij}  c^j}.
\ea
for a symmetric ($S$) and antisymmetric ($A$) matrix  respectively,
to
localize (\ref{fpl}) onto (\ref{fdh}):
$$
\Z ~=~ \int dz^i dc^i  \frac{\sqrt{\det|| \Omega_{ij} ||}}{\sqrt{\det
||g_{ij}||} }~\delta (\X )\delta (c)~ e^{-\phi H - \omega} ~ =~
\int dz^i \frac{\sqrt{\det ||\Omega_{ij} ||}}{\sqrt{\det
||g_{ij} ||}}~\delta (\X)~e^{-\phi H}
$$
$$
=~ \int d\X^i \det || \left(\frac{\partial z^i}{\partial \X
^j}\right ) || \cdot \frac{\sqrt{\det || \Omega_{ij} ||}}{\sqrt{\det
||g_{ij}||}}\delta
(\X)~e^{-\phi H}
$$

$$
=~ \sum _{dH=0} \frac{1}{\det || \partial _i \X ^j ||}
\frac{\sqrt{\det || \Omega_{ij} ||}}{\sqrt{\det || g_{ij}
||}}~e^{-\phi
H}
$$
Since $dH=0$, we have $\X=0$ and
$$
\Omega _{ij} =\partial_i (g_{jk} \X^k) - \partial_j (g_{ik}\X
^k) = g_{jk} \partial_i \X^k - g_{ik} \partial_{j} \X^k
$$
On the other hand, in terms of its components $\cL_{H}g=0$ becomes
$$
\X^k \partial_k g_{ij} + g_{ik} \partial_j \X^k + g_{jk}
\partial_i \X^k = 0
$$
As a consequence,
$$
g_{ik} \partial_j \X^k + g_{jk} \partial_i \X^k = 0
$$
that is
$$
\Omega_{ij} = 2 g_{jk} \partial_i \X^k
$$
Consequently we get
$$
\Z ~=~ \sum\limits_{dH=0} \frac{1}{\det || \partial_i \X^j || } \cdot
\frac{\sqrt{\det || g_{ij} ||}  \cdot \sqrt{\det || \partial_i \X^k
||} }{\sqrt{\det || g_{ij} ||}}  ~\exp\{ -\phi H \}
$$
\be
=~
\sum\limits_{dH=0} { e^{-\phi H} \over \sqrt{\det ||\partial_i\X^j ||
}}
\ee
Since
$$
\partial_i \X^j ~=~ \omega ^{jk}\partial_i\partial_k H
$$
we get finally get the Duistermaat-Heckman integration formula
(\ref{fdh}),
\be
\Z ~=~ \sum\limits_{dH=0} { \sqrt{\det || \om  ||} \over
\sqrt{\det || \partial_{ij} H || } }~\exp\{ -\phi H\}
\ee
for a Hamiltonian $H$ that generates the action of $U(1)$ on the
phase
space.

\vskip 1.5cm
{\bf 4. Generalizations:}
\vskip 1.0 cm

Our discussion in section 2. suggests, that the
localization techniques to evaluate integrals of the form
(\ref{fpf1}) could be
applied to quite general integrable Hamiltonians: The
existence of a functional $\psi$ which is Lie derived by the
Hamiltonian seems to be generally connected to the concept of
integrability.  Since the integral (\ref{fpf1})  depends on
the equivariant cohomology class determined by $\phi H+\omega$ rather
than its given representative, it is natural to expect that
(\ref{fdh}) is just an example of a much more general phenomenon.
At the moment this is not yet well understood, and thus we shall here
only discuss a few possible ways to generalize
the integration formula (\ref{fdh}).

\vskip 0.4cm
First, we shall consider the general case of a compact nonabelian Lie
group $G$
which acts on $M$, and we are interested in a Hamiltonian $H_2$ which
is a
quadratic Casimir for the nonabelian Lie group generators $T_a$,
\be
{\cal C}_2 ~=~ \eta^{ab}T_a T_b
\ee
\be
\{ T_a , {\cal C}_2 \} ~=~ 0
{}~~~~~{\rm for ~ all ~ }a
\ee
Here  $\eta^{ab}$ is positive definite and nondegenerate. We consider
the path
integral
\be
\Z ~=~ \phi^{n} \int dz^i dc^i \exp \{ - \phi \eta^{ab}T_aT_b  -
\omega \}
\ee
We multiply this by
\be
1 ~=~ \left( { \phi \over \pi } \right)^{\frac{n}{2}} \sqrt{ \det
|| \eta^{ab} || } \int\limits_{-\infty}^{\infty} \! dq_a
\exp\{ -\phi \eta^{ab} q_a q_b \}
\ee
which yields
\be
\Z ~=~ \left( { \phi \over \pi } \right)^{\frac{n}{2}} \phi^n \sqrt{
\det ||
\eta^{ab} || } \int\limits_{-\infty}^{\infty} \! dq_a  \exp\{ -
\phi\eta^{ab} q_a q_b \} \int dz^i dc^i
\exp\{ - 2\phi\eta^{ab} q_a T_b  - \omega \}
\label{fqi}
\ee
Denoting
$$
H_q ~=~ 2 \eta^{ab} q_a T_b
$$
which we can identify as the momentum map (\ref{fmm}) with $\phi^a =
2\eta^{ab} q_b$, we then conclude that the second integral in
(\ref{fqi}) is of
the form (\ref{fef}). Consequently (\ref{fqi}) localizes to
\be
\Z ~=~ \left( { \phi \over \pi } \right)^{\frac{n}{2}}
\sqrt{ \det || \eta^{ab} || } \int\limits_{-\infty}^{\infty} \! dq_a
\exp\{ - \phi\eta^{ab} q_a q_b \}
\sum\limits_{d H_q=0} { \sqrt{ det||\om || } \over \sqrt{ \det ||
\partial_{ij} H_q || } } ~\exp\{ -\phi H_q \}
\ee
Generalizations to higher order Casimirs, and for more
general functionals $H[T_a]$ of the Lie algebra generators can also
been considered \cite{witt}, \cite{omat}.

\vskip 0.4cm

Next, we first consider a (not necessarily integrable) Hamiltonian
$H$ with a
number of conserved quantities $Q_{\alpha}$,
\be
\{ H , Q_{\alpha} \} ~=~ 0
\label{fcc2}
\ee
Using these conserved quantities, we introduce the
following ({\it degenerate!}) symmetric matrix
\be
G_{ij} (Q) ~=~ \sum\limits_{\alpha} \partial_{i}
Q_{\alpha} \partial_{j} Q_{\alpha}
\label{fsm}
\ee
We then consider the Lie derivative $\cL_H G$. In components, this
gives
\be
\partial_{i}( \{ H , Q_{\alpha} \} ) \partial_{j} Q_{\alpha} +
\partial_{j} ( \{ H , Q_{\alpha} \} ) \partial_{i} Q_{\alpha}
{}~=~ 0
\ee
as a consequence of (\ref{fcc2}), so that the matrix (\ref{fsm})
satisfies the
Lie derivative condition (\ref{frm}). However, for the same reason we
find
that the corresponding one-form (\ref{flf})
\be
\psi_Q ~=~ G_{ij}\X^{i} c^i ~=~ \omega^{ik} \partial_{k}H
\partial_{i} Q_{\alpha} \partial_{j} Q_{\alpha} c^j ~=~
\{ H , Q_{\alpha} \} dQ_{\alpha} ~=~ 0
\ee
and consequently we do not get a localization formula.

In the loop space (\ref{fsm}) can be used to derive localization
formulas
\cite{omat}, and it  would be interesting to see if a proper variant
of the
present  could be used to derive a partial localization also for the
pertinent
integral (\ref{fpf1}). With $Q_{\alpha}$ the full set of conserved
quantities
for an integrable model $(H,\omega)$ this could then yield a
localization
formula for a quite general integrable system.

\vskip 0.4cm

We shall conclude this article by investigating
properties of the following \cite{topi} alternative geometrical
condition to
the Lie derivative condition (\ref{frm}): We shall assume that
instead  of
(\ref{frm}) we
have a metric tensor which satisfies
\be
\nabla_{\X_H}\X_H ~=~0
\ee
or in components
\be
\X^l \partial_l\X^k + \Gamma^k_{ij} \X^i \X^j = 0
\label{fcf}
\ee
for the Hamiltonian vector field
\be
\X_H^i ~=~ \omega^{ij} \partial_{j} H
\ee
This condition means, that the Hamiltonian flow of $H$ is geodetic to
$g_{ij}$.

We introduce the following Hamiltonian
\be
K ~=~ \half g_{ij} \X_H^i \X_H^j
\ee
and the following symplectic two-form
\be
\Omega_{ij} ~=~ \partial_i (g_{jk}\X^k) - \partial_j (g_{ik}\X^k)
\ee
and argue, that $(H, \omega)$ and $(K, \Omega)$ determines a
bi-hamiltonian pair,  {\it i.e.}
\be
\partial_i K ~=~ \Omega_{ij} \X_H^j
\label{ftu}
\ee

In order to establish this we consider the {\it r.h.s.} of
(\ref{ftu}),
$$
\Omega_{ij}\X^j = [\partial_i ( g_{jk} \X^k ) - \partial_j ( g_{ik}
\X^k ) ]\X^j
$$
\be
=~ (\partial_i g_{ij} ) \X^j \X^k + g_{jk} ( \partial_i \X^k ) \X^j -
(\partial_j g_{ik} ) \X^k \X^j - g_{ik} ( \partial_j \X^k) \X^j
\label{fuy}
\ee
We then use the component form (\ref{ftu}), {\it i.e.}
$$
\X^l ( \partial_l \X^k ) + \frac{1}{2} g^{kl} (\partial_i
g_{jl} + \partial_j g_{il} - \partial_l g_{ij} ) \X^i
\X^j = 0
$$
from which we get
$$
g_{kl} \X^j \partial_j \X^l = - ( \partial_i g_{jk} ) \X^i \X^j +
\frac{1}{2}
(\partial_k g_{ij} ) \X^i \X^j
$$
Substituting this into then last term in (\ref{fuy}), we then get
$$
= ( \partial_i g_{jk} ) \X^j \X^k + g_{jk} (\partial_i \X^k) \X^j -
(\partial_j
g_{ik}) \X^k \X^j + (\partial_k g_{ij}) \X^k \X^j - \frac{1}{2}
(\partial_i
g_{jk}) \X^k \X^j
$$
$$
=~ \frac{1}{2} (\partial_i g_{jk}) \X^j \X^k + g_{jk} (\partial_i
\X^k) \X^j ~
=~ \frac{1}{2} (\partial_i g_{jk}) \X^j \X^k + \frac{1}{2} g_{jk}
(\partial_i
\X^j) \X^k + \frac{1}{2} g_{jk} \X^j \partial_i
\X^k
$$
\be
=~ \partial_i (\frac{1}{2} g_{jk} \X^j \X^k)
\ee
This coincides with the {\it l.h.s.} of (\ref{ftu}), and establishes
the
bi-hamiltonian structure. In fact, we have here shown that
\be
d_{H} (K+\Omega) ~=~ 0 ~~\Leftrightarrow ~~ \nabla_{\X_H}\X_H ~=~0
\label{fue}
\ee

Unfortunately, we also find that the symplectic two-form $\Omega$ is
{\it
degenerate} on a compact phase space \cite{mikk}: Consider the
$\Omega$ -
volume form
\be
\Omega^n ~=~ \sqrt{ det || \Omega_{ij} || } dz^1 \wedge ~ ... ~
\wedge dz^{2n}
\ee
Since
\be
\Omega ~=~ d(i_H g)
\ee
globally on $M$ (in obvious notation), Stokes theorem yields
\be
\int_{M} \Omega^n ~=~
\int_{M} d( i_H g \wedge \Omega^{n-1}) ~=~ 0
\ee
which implies that $det || \Omega_{ij} || = 0$. However, for
integrability it appears to be sufficient, that degeneracies of
$\Omega$ which
occur only on  submanifolds of $M$ with co-dimensions two or more are
not
necessarily fatal, provided the Hamiltonian vanish at these
submanifolds to
keep the equations of motion non-singular.

Finally,  we argue that $K + \Omega$ is equivariantly
exact, {\it i.e.} that there exist a one-form $\psi$ such that
\be
d_H \psi ~=~ K + \Omega
\label{feh1}
\ee
For this, we first observe that if we denote
$$
\eta = i_Hg = g_{ab} \X_H^a c^b
$$
we get
\be
d_H \eta ~=~ 2K + \Omega
\label{feh2}
\ee
If we then subtract (\ref{feh1}) from (\ref{feh2})
and define
$$
\theta = \eta - \psi
$$
we get
\be
d_H \theta ~=~ d_H (\eta - \psi) ~=~ K
\ee
that is
\be
d\theta ~=~ 0
\ee
and
\be
i_H \theta ~=~ K
\ee
If we then assume that $\theta$ is exact so that we can write $\theta
= dF$ for some function $F$ (by Poincare's lemma such a $F$ exists
{\it at least} locally), we obtain
\be
i_H dF ~=~ \X_H^a\partial_a F ~=~ K
\ee
so that
\be
\psi ~=~ i_H g ~-~ dF
\ee
is the desired one-form that satisfies (\ref{feh1}).  Consequently we
have
established that the condition (\ref{fue}) yields relations analogous
to
those that yield the Duistermaat-Heckman formula,  and it would be
interesting
to see how (\ref{fue}) could be used to derive new localization
formulas.

\vskip 1.5cm
{\bf 5. Conclusions}
\vskip 1.0cm

In conclusion, we have investigated the relations between equivariant
cohomology and classical integrability. In particular, we have
explained in
detail how the localization formulas for the classical partition
function
(\ref{fpf1}) are derived from the formalism of equivariant
cohomology, and as
an example we have derived the Duistermaat-Heckman integration
formula in
detail. We have also discussed some generalizations of the
equivariance
structure which underlies the Duistermaat-Heckman integration
formula, and it
would be interesting to see if these generalizations yield new
integration formulas. In particular, it would be very interesting to
understand
fully the relation between integrability and equivariant cohomology,
and
whether localization techniques could in fact be extended to evaluate
the
partition functions for quite generic {\it quantum} integrable
models.

\vskip 1.0cm
We thank M.Laine, K.Palo and O.Tirkkonen for discussions.

\vfill\eject

\end{document}